\documentclass[apj]{emulateapj} 

\usepackage{epsfig}
\usepackage{amsmath}
\usepackage{amssymb}
\usepackage{natbib}
\usepackage{threeparttable} 
\usepackage{booktabs}

\usepackage{epstopdf}

\newcommand{\chan}{\textit{Chandra}}
\newcommand{\swift}{\textit{Swift}}

\newcommand{\xmm}{\textit{XMM-Newton}}

\newcommand{\exosat}{\textit{EXOSAT}}
\newcommand{\rosat}{\textit{ROSAT}}
\newcommand{\einstein}{\textit{Einstein}}

\newcommand{\Msun}{\mathrm{M}_{\odot}}
\newcommand{\lum}{\mathrm{erg~s}^{-1}}
\newcommand{\flux}{\mathrm{erg~cm}^{-2}~\mathrm{s}^{-1}}

\newcommand{\cnts}{\mathrm{counts~s}^{-1}}

\newcommand{\nh}{\mathrm{cm}^{-2}}
\newcommand{\col}{\mathrm{g~cm}^{-2}}

\newcommand{\exo}{EXO 0748--676}
\newcommand{\igr}{IGR J17480--2446}

\newcommand{\sax}{SAX J1750.8--2900}

\newcommand{\swiftpulsar}{Swift J1749.4--2807}

\newcommand{\xte}{XTE J1701--462}

\newcommand{\ks}{KS~1731--260}
\newcommand{\kstwee}{KS~1741--293}
\newcommand{\grs}{GRS~1741.9--2853}
\newcommand{\mxb}{MXB~1659--29}

\newcommand{\maxisource}{MAXI~J0556--332}

\newcommand{\exoter}{EXO 1745--248}

\newcommand{\saxamxp}{SAX J1808.4--3658}

\def \atel {\textit{ATel}}

\slugcomment{Received 2014 March 9; accepted 2014 June 26; published 2014 ???}

\shorttitle{Crust cooling of \exo}
\shortauthors{Degenaar et al.}

\begin{document}

\title{Probing the Crust of the Neutron Star in \exo}

\author{N. Degenaar$^{1,}$\altaffilmark{11}, Z. Medin$^{2}$, A. Cumming$^{3}$, R. Wijnands$^{4}$, M.T.~Wolff$^{5}$, E.M.~Cackett$^6$, J.M.~Miller$^{1}$, P.G.~Jonker$^{7,8}$, J.~Homan$^{9}$, and E.F.~Brown$^{10}$
}
\affil{$^1$Department of Astronomy, University of Michigan, 500 Church Street, Ann Arbor, MI 48109, USA; degenaar@umich.edu\\
$^{2}$Los Alamos National Laboratory, Los Alamos, NM 87545, USA\\
$^3$Department of Physics, McGill University, 3600 rue University, Montreal, QC, H3A 2T8, Canada\\
$^4$Astronomical Institute Anton Pannekoek, University of Amsterdam, Postbus 94249, 1090 GE Amsterdam, The Netherlands\\
$^5$Space Science Division, Naval Research Laboratory, Washington, DC 20375, USA\\
$^6$Department of Physics and Astronomy, Wayne State University, 666 West Hancock Street, Detroit, MI 48201, USA\\
$^7$SRON, Netherlands Institute for Space Research, Sorbonnelaan 2, 3584 CA, Utrecht, The Netherlands\\
$^8$Harvard-Smithsonian  Center for Astrophysics, 60 Garden Street, Cambridge, MA~02138, U.S.A.\\
$^9$MIT Kavli Institute for Astrophysics and Space Research, 77 Massachusetts Avenue, Cambridge, MA 02139, USA\\
$^{10}$Department of Physics and Astronomy, Michigan State University, East Lansing, MI 48824, USA
}
\altaffiltext{11}{Hubble Fellow}

\begin{abstract}
X-ray observations of quiescent X-ray binaries have the potential to provide insight into the structure and the composition of neutron stars. \exo\ had been actively accreting for over 24 yr before its outburst ceased in late 2008. Subsequent X-ray monitoring revealed a gradual decay of the quiescent thermal emission that can be attributed to cooling of the accretion-heated neutron star crust. In this work, we report on new \chan\ and \swift\ observations that extend the quiescent monitoring to $\simeq$5~yr post-outburst. We find that the neutron star temperature remained at $\simeq$117~eV between 2009 and 2011, but had decreased to $\simeq$110~eV in 2013. This suggests that the crust has not fully cooled yet, which is supported by the lower temperature  ($\simeq$95~eV) measured $\simeq$4~yr prior to the accretion phase in 1980. Comparing the data to thermal evolution simulations reveals that the apparent lack of cooling between 2009 and 2011 could possibly be a signature of convection driven by phase separation of light and heavy nuclei in the outer layers of the neutron star. 
\end{abstract}

\keywords{accretion, accretion disks --- binaries: eclipsing --- stars: individual (\exo) --- stars: neutron --- X-rays: binaries}

\section{Introduction}
Transient neutron star low-mass X-ray binaries (LMXBs) are excellent laboratories for increasing our understanding of the structure and the composition of neutron stars, and how matter behaves under extreme physical conditions. In these binary star systems a neutron star is accompanied by a late-type star that overflows its Roche lobe and transfers matter to an accretion disk. This matter is rapidly accreted onto the neutron star during outburst episodes, whereas little or no matter reaches the compact primary during quiescent intervals.

These accretion cycles have a profound effect on the interior properties of neutron stars. They cool during quiescence as they lose thermal energy via photons emitted from their surface and neutrinos escaping from their crust and core \citep[e.g.,][]{yakovlev03,page2006,steiner2009,schatz2014}. However, neutron stars can re-gain thermal energy during accretion outbursts. 

The accretion of matter compresses the crust of a neutron star, which causes successive electron captures, neutron emissions and pycno-nuclear fusion reactions. Together, these processes deposit an energy of $\simeq$2~MeV per accreted nucleon in the crustal layers \citep[e.g.,][]{haensel1990b,haensel1990a,gupta07,steiner2012}. This energy is thermally conducted both towards the stellar core and surface, and can effectively maintain the interior temperature of the neutron star at $\simeq10^{7}-10^{8}$~K. This temperature is set by the energy injected during its historic accretion activity and the efficiency of the neutrino cooling processes \citep[e.g.,][]{brown1998,colpi2001,yakovlev03,wijnands2012}. 

During quiescent episodes, thermal X-rays from the surface of the neutron star may be detected. This allows a measurement of its temperature, which can encode valuable information about its interior properties. 
Of particular interest are observations obtained shortly after the cessation of an outburst; heating due to accretion may lift the temperature of the crust well above that of the stellar core and the subsequent cooling may be observable once back in quiescence \citep[][]{wijnands2001,ushomirsky2001,rutledge2002}. 

Indeed, dedicated X-ray monitoring of six LMXBs (\ks, \mxb, \xte, \exo, \igr, and \maxisource), revealed that the temperature of the neutron star decreased for years following the cessation of accretion, consistent with the heating/cooling paradigm \citep[e.g.,][J. Homan et al., in preparation]{wijnands2002,wijnands2004,cackett2006,cackett2008,cackett2010,degenaar09_exo1,degenaar2011_terzan5_3,degenaar2010_exo2,degenaar2013_ter5,diaztrigo2011,fridriksson2010,fridriksson2011}. Comparison with thermal evolution simulations has yielded valuable insight into the thermal and transport properties of neutron star crusts \citep[][]{shternin07,brown08,degenaar2011_terzan5_3,page2013,turlione2013}.

Despite these successes, interpretation of the quiescent thermal emission and crustal cooling is complicated by the question whether accretion onto the neutron star fully comes to a halt. Searching for (strong) non-thermal emission in the quiescent X-ray spectrum, irregular quiescent X-ray variability, or optical/UV signatures of the quiescent accretion stream can shed light on whether residual accretion occurs \citep[see e.g.,][for recent studies]{cackett2010_cenx4,cackett2011,cackett2013_cenx4,degenaar2012_1745,bernardini2013}.


\subsection{\exo}
The neutron star LMXB \exo\ was discovered almost three decades ago \citep[][]{parmar1985}. The detection of X-ray eclipses indicates that the binary is viewed at high inclination ($i\simeq75^{\circ}-83^{\circ}$), and led to a measurement of the orbital period \citep[$\simeq3.82$~hr;][]{parmar1986,wolff2008c}. The source displays thermonuclear X-ray bursts, which allows for a distance determination \citep[$\simeq7$~kpc; e.g.,][]{galloway2008}.

\exo\ was first detected in outburst in 1984 with \exosat\ \citep[][]{reynolds1999}, and was serendipitously detected in quiescence with \einstein\ in 1980 \citep[][]{parmar1986,garcia1999}. The source remained in outburst for $\simeq$24~yr and during this time the flux was moderately stable with occasional excisions to higher and lower fluxes. However, its activity suddenly ceased in 2008 September \citep[][]{wolff2008,wolff2008b,hynes2008,torres2008}. Subsequent monitoring with \swift, \chan\ and \xmm\ revealed a relatively hot neutron star that gradually cooled over time \citep[][]{degenaar09_exo1,degenaar2010_exo2,diaztrigo2011}. 

In this work we report on new X-ray observations of \exo\ to further monitor how the accretion-heated crust cools, and to search for signs of continued low-level accretion. We also re-analyze the \einstein\ data obtained in 1980 to measure the pre-outburst temperature of the neutron star. We then compare the entire data set to crust cooling simulations.


\section{Observations and Data Analysis}
Table~\ref{tab:obs} gives an overview of all new \chan\ and \swift\ data of \exo. A list of earlier X-ray observations obtained during the quiescent phase can be found in \citet{degenaar09_exo1}, \citet{degenaar2010_exo2} and \citet{diaztrigo2011}. To benefit from the latest calibration updates and to ensure a homogeneous analysis, these quiescent X-ray observations were re-reduced and re-analyzed in this work.

\begin{table}
\begin{center}
\caption{Log of New X-Ray Observations.\label{tab:obs}}
\begin{tabular*}{0.49\textwidth}{@{\extracolsep{\fill}}ccccc}
\hline\hline
Instr. & ObsID & Date & Exp. & Rate \\
& &  & (ks) & ($10^{-2}~\cnts$) \\
\hline
\swift\ & 90420001 & 2010 May 1 & 8.5 & $2.8 \pm0.2$  \\ 
\swift\ & 90420002 & 2010 Jun 22 & 7.7 & $3.0 \pm 0.2$  \\ 
\swift\ & 90420003 & 2010 Aug 22 & 6.5 & $2.7 \pm 0.2$  \\ 
\swift\ & 90420004 & 2010 Aug 24 & 4.0 &	$3.2 \pm 0.3$ \\ 
\chan\ & 11060 & 2010 Oct 20 & 27.2 & $14 \pm 2$ \\
\swift\ & 90420005 & 2010 Oct 21 & 10.3 & $2.9 \pm 0.2$ \\ 
\swift\ & 90420006 & 2010 Dec 23 & 4.0 &$2.8 \pm 0.3$ \\ 
\swift\ & 90420007 & 2010 Dec 24 & 6.3 & $2.1 \pm 0.2$  \\ 
\swift\ & 90420008 & 2011 Feb 22 & 10.5 & $2.5 \pm 0.2$ \\ 
\swift\ & 31272051 & 2011 Jun 28 & 6.0 & $ 2.3\pm 0.3$ \\ 
\swift\ & 31272052 & 2011 Jun 29 & 4.0 & $ 2.7\pm 0.2$ \\ 
\chan\ & 12414 & 2011 Jul 2/3 & 38.1 & $14 \pm 2$ \\
\swift\ & 31272053 & 2013 Mar 23 & 7.8 & $ 2.6\pm0.2$ \\ 
\chan\ & 14663 & 2013 Aug 1 & 42.9 & $9 \pm 1$ \\
\hline
\end{tabular*}
\tablecomments{The listed exposure times and count rates are not corrected for eclipses. The count rates refer to the 0.3--10 keV energy band. Errors represent 90\% confidence levels.
}
\end{center}
\end{table}


\subsection{New \chan\ Observations}\label{subsec:chan}
We obtained three new \chan\ observations of \exo\ between 2010 October and 2013 August (Table~\ref{tab:obs}). The setup was similar to previous quiescent observations of the source, using the S3 chip of the Advanced CCD Imaging Spectrometer \citep[ACIS;][]{garmire2003_acis}. The ACIS-S3 CCD was operated in a 1/8 sub-array during the 2010 observation and in a 1/4 sub-array during the 2011/2013 observations. We reduced the data using the \textsc{ciao} package (ver. 4.5). All observations were free from background flaring. 

We extracted source events using a circular region of $3''$ radius centered on \exo, and a $10''-25''$ annulus was used to estimate the background (excluding a circular region with a $2''$ radius centered on a faint point source). Count rates and light curves were extracted using the task \textsc{dmextract}. Source and background spectra, as well as the corresponding response files, were created using the meta task \textsc{specextract}. We used \textsc{grppha} to group the spectra into bins of at least 20 photons. About 3800--6400 net source events were collected for each \chan\ observation. 

The light curves obtained from the individual observations clearly show the presence of eclipses at times consistent with the ephemeris of \cite{wolff2008c}. During the eclipses, the X-ray emission from the neutron star is temporarily blocked by the companion. To calculate the time-averaged X-ray fluxes excluding the eclipsed epochs, we therefore reduced the exposure times of all \chan\ observations by 500~s per eclipse \citep[which is the observed duration of the eclipses in quiescence;][]{bassa09}.

\begin{table*}
\begin{center}
\caption{Results from Analysis of the Spectral Data\label{tab:spec}}
\begin{tabular*}{1.0\textwidth}{@{\extracolsep{\fill}}cccccccccccc}
\hline\hline
Instrument & Epoch & MJD & $N_{\mathrm{H}}$ & $kT^{\infty}$  & $F_{\mathrm{X}}$ & $F_{\mathrm{X,th}}$  & $F_{\mathrm{bol,th}}$ & $f_{\mathrm{th}}$ & $L_{\mathrm{X}}$ & $L_{\mathrm{bol,th}}$ \\
& & & ($10^{21}~\nh$) & (eV) & \multicolumn{3}{c}{($10^{-12}~\flux)$} & & \multicolumn{2}{c}{($10^{33}~\lum)$} \\
\hline
\einstein & 1980 May & 44381 & $0.74$ fix & $94.6^{+5.6}_{-16.0}$ & $0.25\pm0.13$ & $0.25\pm0.13$ & $0.38\pm0.20$ & $1.00^{+0}_{-0.26}$ & $1.5\pm0.8$ & $ 2.3\pm1.2$ \\
\chan & 2008 Oct & 54755.5 & $1.03 \pm 0.10$ & $129.1 \pm 2.3$ & $1.23 \pm 0.05$ & $1.08 \pm 0.04$ & $1.26 \pm 0.05$ & $0.87\pm 0.06$  & $7.4 \pm 0.3$ & $ 7.6\pm 0.3$ \\
{\it XMM} & 2008 Nov & 54776 & $ 0.51\pm 0.05$ & $126.1 \pm 2.2$ & $1.02 \pm 0.03$ & $0.96 \pm 0.02$ & $1.24 \pm 0.02$ & $0.94\pm 0.03$  & $6.2 \pm 0.2$ & $ 7.5\pm 0.1$ \\
\chan & 2009 Feb & 54886 & $0.93 \pm 0.13$ & $122.6 \pm 2.6$ & $0.89 \pm 0.08$ & $0.85 \pm 0.05$ & $1.02 \pm 0.06$ & $0.96^{+0.04}_{-0.09}$  & $5.4 \pm 0.5$ & $ 6.1\pm 0.4$ \\
{\it XMM} & 2009 Mar & 54908 & $0.51 \pm 0.04$ & $120.0 \pm 2.0$ & $0.77 \pm 0.02$ & $0.77 \pm 0.01$ & $1.02 \pm 0.02$ & $1.00^{+0}_{-0.03}$  & $4.7 \pm 0.1$ & $ 6.1\pm 0.1$ \\
\chan & 2009 Jun & 54992 & $0.78 \pm 0.13$ & $117.8 \pm 2.5$ & $0.75 \pm 0.07$ & $0.70 \pm 0.04$ & $0.85 \pm 0.05$ & $0.94^{+0.06}_{-0.10}$  & $4.5 \pm 0.4$ & $ 5.1\pm 0.3$ \\
{\it XMM} & 2009 Jul & 55013 & $0.42 \pm 0.04$ & $115.5 \pm 2.2$ & $0.72 \pm 0.03$ & $0.65 \pm 0.01$ & $0.87 \pm 0.02$ & $0.90\pm 0.04$  & $4.4 \pm 0.2$ & $ 5.2\pm 0.1$ \\
\chan & 2010 Apr & 55306 & $0.86 \pm 0.14$ & $116.8 \pm 2.5$ & $0.76 \pm 0.07$ & $0.68 \pm 0.04$ & $0.82 \pm 0.05$ & $0.89\pm 0.11$  & $4.6 \pm 0.4$ & $ 5.0\pm 0.3$ \\
{\it XMM} & 2010 Jun & 55364 & $0.54 \pm 0.04$ & $116.2 \pm 1.9$ & $0.67 \pm 0.02$ & $0.67 \pm 0.01$ & $0.89 \pm 0.02$ & $1.00^{+0}_{-0.02}$  & $4.0 \pm 0.1$ & $ 5.4\pm 0.1$ \\
\chan & 2010 Oct & 55489 & $0.64 \pm 0.13$ & $115.4 \pm 2.2$ & $0.68 \pm 0.04$ & $0.64 \pm 0.03$ & $0.78 \pm 0.04$ & $0.95^{+0.05}_{-0.07}$  & $4.1 \pm 0.2$ & $ 4.7\pm 0.2$ \\
\chan & 2011 Jul & 55745 & $0.95 \pm 0.12$ & $117.6 \pm 2.2$ & $0.80 \pm 0.04$ & $0.70 \pm 0.03$ & $0.85 \pm 0.04$ & $0.87\pm 0.06$  & $4.8 \pm 0.2$ & $ 5.1\pm 0.2$ \\
\chan & 2013 Aug & 56505 & $0.91 \pm 0.15$ & $109.9 \pm 2.0$ & $0.54 \pm 0.03$ & $0.51 \pm 0.03$ & $0.63 \pm 0.03$ & $0.96^{+0.04}_{-0.07}$  & $3.2 \pm 0.2$ & $ 3.8\pm 0.2$ \\
\hline
\end{tabular*}
\tablecomments{Quoted uncertainties represent 90\% confidence intervals. $F_{\mathrm{X}}$ represents the total unabsorbed model flux and $F_{\mathrm{X,th}}$ gives the thermal flux (0.5--10 keV). The ratio of these two fluxes, $f_{\mathrm{th}}$, represents the fractional contribution of the thermal emission to the total model flux. $F_{\mathrm{bol,th}}$ represents the unabsorbed thermal flux in the 0.01--100 keV energy band. The 0.5--10 keV total model luminosity ($L_{\mathrm{X}}$) and the thermal bolometric luminosity ($L_{\mathrm{bol,th}}$) are for a distance of $D=7.1$~kpc. 
} 
\end{center}
\end{table*}


\subsection{New \swift\ Observations}\label{subsec:swift}
\exo\ has been observed with the X-Ray Telescope \citep[XRT;][]{burrows05} onboard \swift\ many times since it transitioned to quiescence \citep[see][]{wolff2008,wolff2008b,degenaar09_exo1,degenaar2010_exo2}. Here we report on 11 new observations that were carried out between 2010 May and 2013 March (see Table~\ref{tab:obs}). All XRT data were obtained in the photon-counting mode. 

The \swift\ observations were reduced using the \textsc{heasoft} suite (ver. 6.13). We first processed the raw XRT data using the \textsc{xrtpipeline}. Employing \textsc{xselect}, we then extracted source events using a circular region with a radius of $35''$, which optimized the signal to noise ratio for the observed count rates \citep[cf.][]{evans2007}. A surrounding annular region of radius $100''-200''$ was used as a background reference. We obtained $\simeq$50--350 net source counts per observation. 

The \chan\ and \xmm\ observations provide superior spectral quality compared to the short \swift/XRT pointings. However, the \swift\ data offer unique dense sampling that allows us to closely track intensity variations occurring on a time scale of days--weeks. For the \swift\ data we therefore focused our attention on the long-term quiescent light curve. The count rates extracted for each observation were corrected for losses due to bad pixels and bad columns. Due to the relatively low count rates, it was not possible to identify eclipses in the data. We therefore checked the good time intervals against the ephemeris of \cite{wolff2008c}; if (part of) eclipses were expected, we reduced the exposure times accordingly.


\subsection{Archival \xmm\ Observations}\label{subsec:xmm}
Following its return to quiescence, \exo\ was observed with the European Photon Imaging Camera \citep[EPIC;][]{struder2001_pn,turner2001_mos} onboard \xmm\ on four occasions \citep[][]{bassa09,diaztrigo2011}. We reduced and analyzed these data using the \textsc{sas} package (ver. 13.0). After reprocessing with \textsc{emproc} and \textsc{epproc}, and removing background flaring events, we extracted spectra and light curves using the task \textsc{evselect}. A circular region with a radius of $35''$ was used for the source and a $70''$-radius circular region placed on an empty part of the CCD served as the background. 

Response files were generated using \textsc{arfgen} and \textsc{rmfgen}. The spectra and response files of the three detectors (MOS1, MOS2, and PN) were combined with the tool \textsc{epicspeccombine}.\footnote[12]{For all four \xmm\ observations we found that the fits results for the combined PN/MOS spectra were consistent with fits to the separate PN and MOS spectra.} Using \textsc{grppha} we grouped the spectral data to contain at least 20 photons per bin. The number of net source counts collected for each \xmm\ observation was $\simeq$24\,000--45\,000 (all three EPIC instruments summed). Similar to our treatment of the \chan\ data, the exposure times of the observations were reduced with 500~s per eclipse.


\subsection{Pre-outburst \einstein\ Observation}\label{subsec:einstein}
\exo\ was serendipitously detected with the Imaging Proportional Counter (IPC) onboard the \einstein\ observatory on 1980 May 22 \citep[ObsID 7708;][]{parmar1986,garcia1999}. 
The source intensity during the $\simeq5.8$~ks exposure was $\simeq (1.7\pm0.2)\times10^{-2}~\cnts$, whereas the local background is estimated at $\simeq (0.6\pm0.1)\times10^{-2}~\cnts$. We created spectra and response files using \textsc{XSelect}. Source events were collected from a region of 20 pixels ($2.7'$), and background events using a source-free region of twice that size. Extrapolating the ephemeris of \citet{wolff2008c} back to the time of the \einstein\ observation suggests that the source may have been eclipsed for $\simeq230$~s. We therefore reduced the exposure time by this amount.

There is little reported information in the literature about cross-calibration between the \einstein/IPC and the current generation of X-ray instruments \citep[e.g., the Crab was too bright for the IPC;][]{kirsch2005}. A study of nova-like variables yielded strong consistency with \rosat/PSPC results, lending credence to the low-energy response of the IPC \citep[][]{verbunt1997}. Systematic uncertainties are estimated to be at the 10\%-15\% level \citep[e.g.,][]{fabricant1984,david1990}. To account for (cross-)calibration uncertainties, we included a conservative 20\% systematic error for the \einstein\ data in our spectral fits.


\subsection{Spectral Fitting Procedures}\label{subsec:spectraldata}
The \chan, \xmm\ and \einstein\ spectra (each with their own response and background files) were fitted simultaneously in the 0.3--10 keV energy range using \textsc{xspec} \citep[ver. 12.8;][]{xspec}. The \einstein\ data and the last three \xmm\ observations were all dominated by background noise above an energy of $\simeq 3$~keV \citep[see also][]{diaztrigo2011}. For these observations we therefore excluded the data above 3 keV.

Expanding on previous work \citep[][]{degenaar09_exo1,degenaar2010_exo2,diaztrigo2011}, we concentrated on fitting the data to a combination of a neutron star atmosphere model \citep[\textsc{nsatmos};][]{heinke2006}, and a power-law (\textsc{pegpwrl}; to model any possible non-thermal emission). To account for the neutral hydrogen absorption along the line of sight, $N_{\mathrm{H}}$, we included the \textsc{tbabs} model adopting the \textsc{vern} cross-sections and \textsc{wilm} abundances \citep[][]{verner1996,wilms2000}. 

Since \exo\ is viewed at a relatively high inclination and hence the absorption along the line of sight could be variable \citep[see, e.g.,][]{cackett2013_1659}, we allowed $N_{\mathrm{H}}$ to change between the different observations. Only for the \einstein\ spectrum we could not obtain good constraints on $N_{\mathrm{H}}$. For this data set we therefore fixed $N_{\mathrm{H}}$ to the average value obtained for the \chan\ and \xmm\ spectra (Table~\ref{tab:spec}). 

For the \textsc{nsatmos} model we fixed the normalization at unity, i.e., we assumed that the emitting area was the same during all observations and corresponded to the entire neutron star surface. When fixing the neutron star mass and radius to $M_{\mathrm{NS}}=1.4~\Msun$ and $R_{\mathrm{NS}}=$10~km, no acceptable fit could be obtained \citep[see also][]{degenaar09_exo1,degenaar2010_exo2,diaztrigo2011}. Both parameters were therefore left free to find the best fit values (assuming that $M_{\mathrm{NS}}$ and $R_{\mathrm{NS}}$ did not change over time, they were tied between the different data sets). The source distance was not well constrained when left to vary freely and drove the neutron star mass and radius to unrealistic values. We therefore fixed the distance to $D=7.1$~kpc in all spectral fits \citep[][]{galloway2008}. The neutron star temperature was allowed to vary freely.

For the non-thermal emission component it was not possible to constrain the power-law index for each observation individually. We therefore chose to fix this parameter for the entire data set to the value obtained from the first \xmm\ observation, which provided the best constraints \citep[$\Gamma=1.7$;][]{degenaar2010_exo2}. The power-law normalization was free to vary.

From the fit results we calculated the effective neutron star temperature as seen by an observer at infinity, $kT^{\infty}_{\mathrm{eff}}= kT_{\mathrm{eff}}/(1+z)$. Here, $(1+z) = (1-R_{\mathrm{s}}/R_{\mathrm{NS}})^{-1/2}$ is the gravitational redshift factor, with $R_{\mathrm{s}}=2GM_{\mathrm{NS}}/c^2$ being the Schwarzschild radius, $G$ the gravitational constant and $c$ the speed of light. Using \textsc{cflux} in \textsc{xspec}, we determined the unabsorbed 0.5--10 keV flux and the thermal bolometric flux. The latter was estimated by extrapolating the \textsc{nsatmos} model component to the 0.01--100 keV energy range. All fluxes were converted to luminosities by assuming $D=7.1$~kpc.

\begin{figure}
 \begin{center}     
  \includegraphics[width=8.6cm]{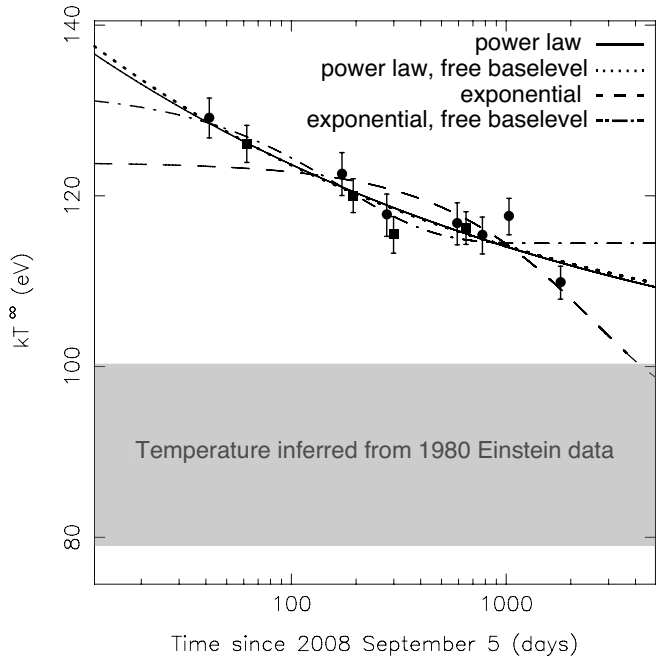}
    \end{center}
\caption[]{{Evolution of the neutron star temperature from \chan\ (filled circles) and \xmm\ (squares) data. Shown are power law (solid) and exponential (dashed) decay fits with a base level set to the pre-outburst temperature measured in 1980 (gray shaded area). The dotted (power law) and dashed-dotted (exponential) lines are decay fits with the quiescent base level left as a free parameter. Error bars indicate 90\% confidence levels.}}
 \label{fig:cool}
\end{figure}

\begin{table}
\begin{center}
\caption{Decay Fits to the Crust Cooling Curve\label{tab:lc}}
\begin{tabular*}{0.4\textwidth}{@{\extracolsep{\fill}}lc}
\hline\hline
Fit Parameter (unit) & Value  \\
\hline
\multicolumn{2}{c}{{ Exponential decay, base level fixed}}  \\
\hline
Normalization, $a$ (eV) & $29.3 \pm  1.1$   \\
Decay time, $\tau$ (days) & $2533.8 \pm  458.1$  \\
Constant offset, $b$ (eV) & 94.6 fixed   \\
$\chi^2_{\nu}$ (dof) & 2.2 (9)   \\
$P_{\chi}$ & 0.02  \\
\hline
\multicolumn{2}{c}{{ Exponential decay, base level free}}  \\
\hline
Normalization, $a$ (eV)  & $17.9 \pm  3.2$   \\
Decay time, $\tau$ (days) & $172.1 \pm  51.7$  \\
Constant offset, $b$ (eV) & $114.4 \pm  1.2$   \\
$\chi^2_{\nu}$ (dof) & 1.1 (10)   \\
$P_{\chi}$ & 0.40  \\
\hline
\multicolumn{2}{c}{{ Power-law decay, base level fixed} } \\
\hline
Normalization, $a$ (eV) & $64.9 \pm  8.4$   \\
Decay index, $\alpha$ & $0.18 \pm  0.02$  \\
Constant offset, $b$ (eV) & 94.6 fixed   \\
$\chi^2_{\nu}$ (dof) & 0.8 (9)   \\
$P_{\chi}$ & 0.58   \\
\hline
\multicolumn{2}{c}{{ Power-law decay, base level free}}  \\
\hline
Normalization, $a$ (eV) & $65.4 \pm  10.2$  \\
Decay index, $\alpha$ & $0.21 \pm  0.03$  \\
Constant offset, $b$ (eV) & $99.1 \pm  2.3$   \\
$\chi^2_{\nu}$ (dof) & 0.8 (10)   \\
$P_{\chi}$ & 0.68 \\
\hline
\multicolumn{2}{c}{{ Power-law decay, no constant offset}}  \\
\hline
Normalization, $a$ (eV) & $146.5 \pm  4.2$  \\
Decay index, $\alpha$ & $0.04 \pm  0.01$  \\
Constant offset, $b$ (eV) & $0$ fixed   \\
$\chi^2_{\nu}$ (dof) & 0.9 (9)   \\
$P_{\chi}$ &  0.54 \\
\hline
\end{tabular*}
\tablecomments{The data was fit to an exponential decay of the form $y(t)=a~e^{-(t-t_0)/\tau}+b$, and a power-law decay shaped as $y(t)=a~(t-t_0)^{-\alpha}+b$. The fixed constant offset of $b=94.6$~eV corresponds to the temperature inferred from the pre-outburst \einstein\ observation. The power-law fit without constant offset is included to allow for a comparison with the literature. In all fits the start of the cooling curve, $t_0$ was set to 2008 September 5 \citep[MJD 54714;][]{degenaar09_exo1}. 
}
\end{center}
\end{table}


\section{Results}

\subsection{X-Ray Spectral Evolution}\label{subsec:evolution}
Simultaneously fitting the \chan, \xmm\ and \einstein\ spectra to a combined neutron star atmosphere and power-law model as described in Section~\ref{subsec:spectraldata}, resulted in a reduced chi-squared value of $\chi^2_{\nu}=1.00$ for 1950 degrees of freedom (dof) with a p-value of $p_{\chi}=0.46$. The best fit yielded a neutron star mass of $M_{\mathrm{NS}}= 1.64 \pm 0.38~\Msun$, and a radius of $R_{\mathrm{NS}} = 13.2^{+0.6}_{-2.0}$~km. The uncertainty in $M_{\mathrm{NS}}$ dominates the errors in all other parameters \citep[see also the discussion in][]{diaztrigo2011}. For the error calculation we therefore fixed $M_{\mathrm{NS}}=1.64~\Msun$ at the best fit value (whereas $R_{\mathrm{NS}}$ was still free). The results are summarized in Table~\ref{tab:spec}.

We find that the values of $N_{\mathrm{H}}$ obtained for the \xmm\ data are systematically lower than for the \chan\ observations. This likely arises due to cross-calibration uncertainties \citep[][]{kirsch2005,degenaar2010_exo2,tsujimoto2011}.\footnote[13]{We note that in \citet{degenaar2010_exo2} it was found that the first \xmm\ observation had an elevated temperature compared to adjacent \chan\ and \swift\ observations. In those spectral fits $N_{\mathrm{H}}$ was fixed between the different data sets. Here we leave $N_{\mathrm{H}}$ free, which does not yield an elevated temperature for the first \xmm\ observation (see Table~\ref{tab:spec} and Figure~\ref{fig:cool}).} Nevertheless, the temperature evolution shows the same trend across the two data sets, indicating that the difference in $N_{\mathrm{H}}$ does not affect the relative temperature changes \citep[see also][]{diaztrigo2011}. There is little variation in $N_{\mathrm{H}}$ among the \chan\ observations, and the same is true for the \xmm\ data set. It therefore appears there are no large changes in the absorption along the line of sight between 2008 and 2013, despite the binary being viewed at high inclination. 

The first \chan\ and first \xmm\ observation (obtained within 2 months after the outburst) required the presence of a hard spectral component, although its contribution to the total 0.5--10 keV flux was small \citep[$\lesssim$15\%;][]{degenaar2010_exo2,diaztrigo2011}. This non-thermal  component is not statistically required in subsequent observations, although it may still account for up to $\simeq$10\% of the 0.5--10 keV flux (Table~\ref{tab:spec}). The quiescent spectra of \exo\ are thus strongly dominated by soft, thermal emission throughout the quiescent phase (see also Section~\ref{subsec:quiescent}).

The neutron star temperature gradually decreased by $\simeq$11~eV within the first year after the outburst (2008 October till 2009 June), but showed little change in the subsequent two years. Indeed, requiring the temperatures of the 6 observations obtained between 2009 June and 2011 July to be the same results in a good fit ($\chi^2_{\nu}=1.01$ for 1955 dof, $p_{\chi}=0.39$), with an average temperature of $kT^{\infty}_{\mathrm{eff}}=116.7\pm1.9$~eV. 
It therefore appeared that the neutron star crust had fully cooled well within two years of entering quiescence \citep[][]{degenaar2010_exo2,diaztrigo2011}. However, our \chan\ observation obtained in 2013 August ($\simeq4.9$ yr after the outburst) shows a decrease in temperature by $\simeq$7~eV compared to 2009--2011 (Table~\ref{tab:spec}). This is illustrated by Figure~\ref{fig:cool}, which shows the evolution of the temperature over time. Forcing the 2013 temperature to be the same as during the previous 6 observations results in a poor fit ($\chi^2_{\nu}=1.07$ for 1956 dof, $p_{\chi}=0.02$).

Since we found little variation in the absorption along the line of sight, the drop in temperature observed in 2013 appears to be genuine, and hence indicates continued or accelerated cooling of the neutron star (see Section~\ref{subsec:decay}). This is supported by the fact that the temperature inferred from the pre-outburst \einstein\ observation ($\simeq$95 eV) is lower than in our latest \chan\ observation (see Table~\ref{tab:spec} and Figure~\ref{fig:cool}). We note that different values of $N_{\mathrm{H}}$, $D$, $M_{\mathrm{NS}}$, and $R_{\mathrm{NS}}$ may shift the absolute temperatures by a few ($\lesssim10$) eV. However, the observed relative temperature change, i.e., the crust cooling curve, is not affected by these systematic uncertainties.

 
\subsection{Crust Cooling Curve Fits}\label{subsec:decay}
To characterize the temperature evolution of \exo\ and compare it with that of other sources, we fitted the crust cooling curve to an exponential decay of the form $y(t)=a~e^{-(t-t_0)/\tau}+b$. Here, $a$ is a normalization, $\tau$ the $e$-folding time, and $b$ represents the quiescent base level. The reference time $t_0$ is set to the presumed end of the outburst, 2008 September 5 \citep[MJD 54714;][]{degenaar09_exo1}. Fixing $b = 94.6$~eV to the 1980 level does not provide an acceptable fit (yielding $\chi^2_{\nu}=2.2$ for 9 dof, $p_{\chi}=0.02$). The fit improves when the base level is allowed to vary ($\chi^2_{\nu}=1.1$ for 10 dof, $p_{\chi}=0.40$), but the obtained value of $b=114.4\pm1.2$~eV is higher than observed with \chan\ in 2013 ($b=109.9\pm2.0$~eV). An exponential decay therefore may not be a good representation of the crust cooling curve of \exo\ (dashed and dashed-dotted lines in Figure~\ref{fig:cool}).

We also fitted the data to a power-law function of the form $y(t)=a(t-t_0)^{-\alpha}+b$, which is the theoretically expected shape for a cooling crust \citep[e.g.,][]{brown08}. Fixing  $b=94.6$~eV to the \einstein\ level yields a good fit with a decay index of $\alpha=0.18\pm 0.02$ ($\chi^2_{\nu}=0.8$ for 9 dof, $p_{\chi}=0.58$). With the constant offset allowed to vary we found $b=99.1\pm2.3$~eV and $\alpha=0.21\pm 0.03$ ($\chi^2_{\nu}=0.8$ for 10 dof, $p_{\chi}=0.68$). The  base level we obtain is consistent within the errors with the \einstein\ measured temperature. The power-law decay fits are indicated by the solid and dotted lines in Figure~\ref{fig:cool}.

The results of the exponential and power-law decay fits are summarized in Table~\ref{tab:lc}. In the literature the crust cooling curves are often fit to a power-law decay without a constant offset. Since the inclusion of a base level changes the resulting slope significantly, we also reference a fit without a constant level to allow for a direct comparison with other sources.

\begin{figure}
 \begin{center}     
  \includegraphics[width=8.7cm, angle=0]{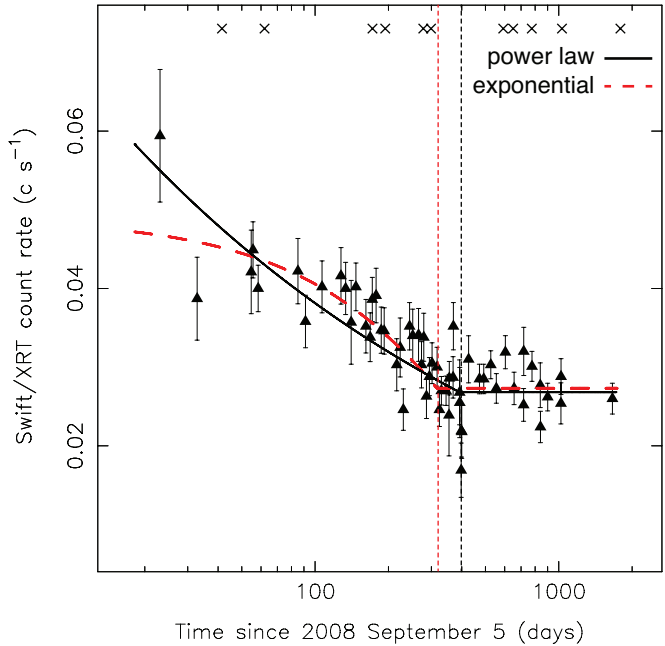}
    \end{center}
\caption[]{{\swift/XRT count rate light curve covering the epoch of 2008--2013 after the source transitioned to quiescence (binned per observation). The black solid and red dashed lines represent  fits to an exponential and a power-law decay, respectively (leveling off to a constant). The black and red dotted lines indicate the times of a transition to a constant level for these fits. The markers on top indicate the times of \chan\ and \xmm\ observations. Error bars represent 90\% confidence intervals.
}}
 \label{fig:swift}
\end{figure}

\subsection{\swift\ Quiescent X-Ray Light Curve}\label{subsec:swiftlc}
The 2008--2013 \swift/XRT count rate light curve of \exo\ is shown in Figure~\ref{fig:swift}. The source intensity changed gradually from $\simeq 6 \times10^{-2}$ to $\simeq 2.5 \times10^{-2}~\cnts$ within the first year after the outburst, but showed little variation thereafter. Simple decay fits with an exponential ($\chi_{\nu}^2=1.5$ for 56 dof) or a power-law function ($\chi_{\nu}^2=1.7$ for 56 dof) leveling off to a constant suggests that the quiescent light curve flattened $\simeq$1~yr post-outburst (Figure~\ref{fig:swift}). This may indicate an episode of relatively constant intensity, as was also hinted by our analysis of the \chan/\xmm\ spectral data (Section~\ref{subsec:evolution}). The \swift\ spectral data is not of sufficient quality to test whether the last data point (obtained in 2013 March) supports the lower temperature seen during the 2013 August \chan\ observation.

There is also more stochastic variability among the data points, although the 90\% error bars largely overlap. Since we corrected the count rates for dead zones on the CCD and the possible occurrence of eclipses (Section~\ref{subsec:swift}), we suspect that the variations are due to photon statistics (this is perhaps supported by the fact that the largest outliers in the light curve concern short observations, i.e., which collected a small number of photons). It is clear that there are no strong intensity flares as are sometimes seen in quiescent neutron star LMXBs (see Section~\ref{subsec:quiescent}).

\subsection{Thermal Evolution Simulations}\label{subsec:sim}
We briefly explored thermal evolution simulations to gauge the properties of the neutron star in \exo. Using the physical model described in \citet{brown08} and \citet{medin2014}, we calculated source-specific cooling curves assuming an outburst duration of $t_{\mathrm{ob}} = $24~yr, and an outburst mass-accretion rate of $\dot{M}_{\mathrm{ob}}= 3\times 10^{16}~\mathrm{g~s}^{-1}$ \citep[e.g.,][]{degenaar2010_exo2}. The model parameters that are then adjusted to match the data are the core temperature $T_{\mathrm{core}}$, and the impurity parameter $Q_{\mathrm{imp}}$. The latter parametrizes how organized the structure of the ion lattice is and hence determines the thermal conductivity \citep[e.g.,][]{itoh1993,brown08}. 

In recent years, evidence has accumulated that there is more heat generated in the crusts of neutron stars than is currently accounted for by nuclear heating models \citep[e.g.,][]{brown08,degenaar2011_terzan5_3,degenaar2013_xtej1709,schatz2014}. We therefore allowed for the inclusion of an additional heat source $Q_{\mathrm{extra}}$ placed at a column depth of $y=1\times10^{12}~\col$. Table~\ref{tab:model} gives an overview of the model calculations. %

\begin{figure}
 \begin{center}     
  \includegraphics[width=8.6cm]{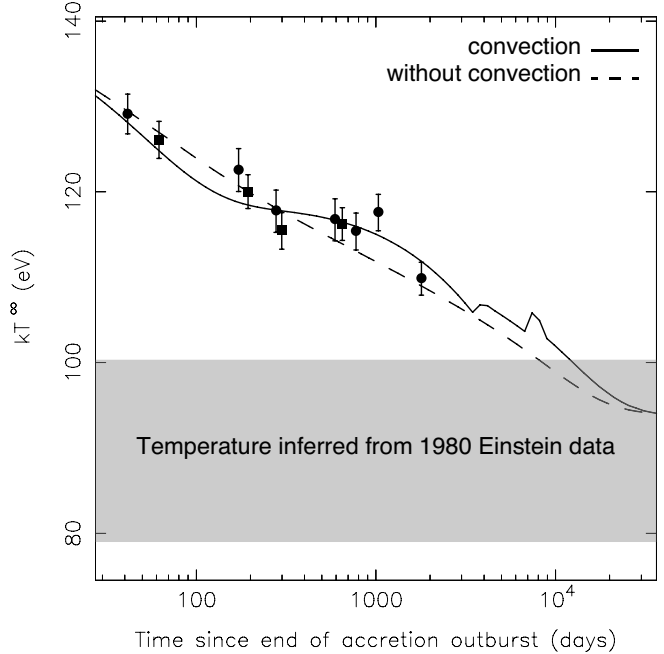}
    \end{center}
\caption[]{{Crust cooling curve of \exo\ constructed from \chan\ (filled circles) and \xmm\ (squares) data. The solid and dashed curves represent model calculations with and without convection taken into account, respectively (see Section~\ref{subsec:sim} for details). The late-time wiggles in the convection curve (near $\simeq$4000 and 9000 days post-outburst) are numerical artifacts. The gray shaded area indicates the pre-outburst temperature. Error bars represent 90\% confidence intervals.}}
 \label{fig:model}
\end{figure}

We find that reproducing the observed high temperatures of \exo\ requires a crust impurity parameter of $Q_{\mathrm{imp}}=40$, and an additional heat source of $Q_{\mathrm{extra}}=1.8$~MeV~nucleon$^{-1}$. This model produces the data reasonably well (dashed curve in Figure~\ref{fig:model}). It suggests that the source has so far moved along a continuous cooling track and will reach its pre-outburst quiescent level many years from now. 

Interestingly, the occurrence of a possible plateau of slow cooling in \exo\ starting $\simeq$1~yr post-outburst (Sections~\ref{subsec:evolution} and~\ref{subsec:swiftlc}) resembles recent calculations of \cite{medin2014}. These authors showed how cooling curves are affected by a convective heat flux that arises when light and heavy nuclei in the crust separate out in liquid and solid phases \citep[][]{horowitz2007,medin2011}. 
Therefore, we also performed model calculations with compositionally-driven convection taken into account. This introduces one extra fit parameter, $X_{\mathrm{b}}$, which is the mass fraction of oxygen at the base of the liquid ocean \citep[for details, see][]{medin2014}. 

We find that the inclusion of convection leads to a plateau of slow cooling between $\simeq$150--750 days post-outburst, broadly consistent with the data (solid curve in Figure~\ref{fig:model}). This arises because the compositionally-driven convection transports heat inward, temporarily slowing the cooling in the crustal layers where the phase separation occurs. We note that the model including convection is not statistically preferred over the non-convective case.

\begin{table}
\begin{center}
\caption{Thermal Evolution Models\label{tab:model}}
\begin{tabular*}{0.49\textwidth}{@{\extracolsep{\fill}}lcc}
\hline\hline
Parameter (unit) & No Convection & Convection  \\
\hline
$Q_{\mathrm{imp}}$ & 40 & 40   \\
$Q_{\mathrm{extra}}$ (MeV~nucleon$^{-1}$) & 1.8 & 1.8    \\
$X_{\mathrm{b}}$ & 0.30 & 0.37   \\
$T_{\mathrm{core}}$ (K)  & $1.50\times10^{8}$ & $1.35\times10^{8}$    \\
\hline
\end{tabular*}
\tablecomments{$Q_{\mathrm{imp}}$ represents the impurity parameter, $Q_{\mathrm{extra}}$ the additional heat energy (placed at a column depth of $y=1\times10^{12}~\col$), $X_{\mathrm{b}}$ the mass fraction of oxygen at the base of the liquid ocean, and $T_{\mathrm{core}}$ the core temperature. We assumed an outburst duration of $t_{\mathrm{ob}} = $24~yr, and mass-accretion rate of $\dot{M}_{\mathrm{ob}}=3\times 10^{16}~\mathrm{g~s}^{-1}$. The equation of state used in these simulations results in $M_{\mathrm{NS}}=1.62~\Msun$ and $R_{\mathrm{NS}}=11.2$~km \citep[for details, see][]{medin2014}.
}
\end{center}
\end{table}


\section{Discussion}\label{sec:discussion}
\subsection{Crustal Cooling in \exo}\label{subsec:cool}
Our new \chan\ and \swift\ observations of \exo\ extend the quiescent monitoring to $\simeq$4.9~yr after the cessation of its very long ($\simeq$24~yr) active period. We find that the neutron star temperature gradually decreased during this time, consistent with expectations for cooling of the accretion-heated neutron star crust. In the first year of quiescence, between 2008 and 2009, the temperature decreased from $kT^{\infty}_{\mathrm{eff}}\simeq129$ to $118$~eV. It then hovered around $117$~eV for at least $\simeq2$~yr till 2011, but our most recent observation obtained in 2013 indicates a further decrease in temperature to $kT^{\infty}_{\mathrm{eff}}\simeq110$~eV. 

Despite the high inclination of the binary, there are no indications that the lower temperature in 2013 is due to a changing absorption column density, such as possibly seen in \mxb\ \citep[][]{cackett2013_1659}. Whereas the apparent lack of temperature evolution after 2009 led to the suggestion that the neutron star crust restored equilibrium with the core \citep[][]{degenaar2010_exo2,diaztrigo2011}, the new data presented in this work suggests that cooling is still ongoing and that a further decrease in temperature may be expected.\footnote[14]{\citet{degenaar2010_exo2} noted that the \einstein\ flux reported by \citet{garcia1999} was consistent within the errors with that inferred from the 2010 \chan\ data. However, \citet{garcia1999} used a different physical model to fit the spectrum, which may introduce biases. Fitting the data in tandem with the post-outburst \chan\ and \xmm\ observations suggests that the pre-outburst temperature was lower than currently seen, provided the caveats mentioned in Section~\ref{subsec:einstein}.} This is supported by the lower temperature measured $\simeq4$~yr prior to the outburst in 1980; $kT^{\infty}_{\mathrm{eff}}\simeq$95~eV.

\subsection{A Signature of Convection?}\label{subsec:convection}
The possible ``plateau'' of stalled cooling starting $\simeq1$~yr post-outburst is reminiscent of the crust cooling curve of \xte. That source too appeared to level off within $\simeq$2 yr of entering quiescence \citep[][]{fridriksson2011}, but \citet{page2013} predicted that after a temporary plateau an accelerated temperature decay would occur, which seems to be borne out by more recent observations (J. K. Fridriksson et al., in preparation). \xte\ experienced a relatively short ($\simeq$1.6 yr) but very bright accretion phase (an average flux near the Eddington limit). As a result of this vigorous heating, the temperature in the crust likely did not reach a steady state but rather had double peaked profile, which would naturally give rise a plateau. This is in sharp contrast to \exo, which was active for 24 yr at relatively low X-ray flux ($\simeq$5\% of Eddington), implying that the crust had ample time to reach a steady state profile \citep[cf.][]{brown08}.  

Another mechanism that may cause a plateau in the cooling curve is a convective heat flux driven by chemical separation of light and heavy nuclei in the outer layers of the neutron star \citep[][]{horowitz2007,medin2011,medin2014}. Inclusion of the inward heat transport by compositionally-driven convection in the model calculations for \exo\ leads to an episode of slow cooling that is broadly consistent with the observations. The crust cooling curve of \exo\ may thus bear an imprint of this process, although the data can also be satisfactory modeled without convection. Perhaps another possibility is that a crustal shell of rapid neutrino cooling as recently identified by \citet{schatz2014} is connected to the period of stalled cooling. This process is highly temperature-sensitive and may be related to the fact that a plateau appears to be seen only in the two hottest crust-cooling neutron stars \exo\ and \xte. However, this could also be an observational bias, since these two sources were more intensely monitored than the others \citep[see][for a comparison]{degenaar2010_exo2}. Further theoretical investigation is required to grasp the implications of this newly identified cooling process on neutron star crust cooling curves. 

It is of note that the model calculations of \exo\ require rather high values for the impurity parameter ($Q_{\mathrm{imp}}=40$), and the additional heat ($Q_{\mathrm{extra}}=1.8$~MeV~nucleon$^{-1}$), to keep the crust hot as long as observed. In contrast, the crust cooling curves of \ks, \mxb, \xte, and \igr\ suggested an impurity parameter of order unity \citep[][]{brown08,degenaar2011_terzan5_3,page2013,medin2014}, consistent with expectations from molecular dynamics simulations \citep[][]{horowitz2008}. In fact, taking into account allowed ranges in mass, radius, and accretion rate, \citet{brown08} set an upper limit of $Q_{\mathrm{imp}}\lesssim 10$ for \ks\ and \mxb. The higher value that we find here could imply that the crust of \exo\ has a more impure (i.e., less organized) structure than the other neutron stars, although it is unclear why that would be the case. Moreover, the obtained value of $Q_{\mathrm{imp}}$ is sensitive to other model parameters.

If we allow for a higher mass-accretion rate, e.g., $\dot{M}= 1.2\times 10^{17}~\mathrm{g~s}^{-1}$, the crust temperature rises and therefore we require a lower impurity parameter ($Q_{\mathrm{imp}}=20$), and less extra heat ($Q_{\mathrm{extra}}=0.35$~MeV~nucleon$^{-1}$). This mass-accretion rate is higher than inferred from X-ray observations \citep[$\dot{M}\simeq3\times 10^{16}~\mathrm{g~s}^{-1}$; e.g.,][]{degenaar2010_exo2,diaztrigo2011}, but not implausible. There are large uncertainties in determining the accretion rate from X-ray observations, in particular for high-inclination systems such as \exo\ when part of the central X-ray source may be blocked from our line of sight, causing $\dot{M}$ to be underestimated. Nevertheless, even for this higher accretion rate $Q_{\mathrm{imp}}$ remains considerably larger than found for the other sources.  Another possible way of keeping the crust in \exo\ hot for a long time is residual accretion during quiescence.

\subsection{On the Possibility of Quiescent Accretion}\label{subsec:quiescent}
Our interpretation of the observations of \exo\ in the crustal heating/cooling framework relies on the assumption that accretion onto the neutron star stopped when the source transitioned to quiescence. It is not straightforward to test this hypothesis with observations. Low-level accretion may generate a thermal emission spectrum like that of a cooling neutron star \citep[][]{zampieri1995,soria2011}. However, the measured temperature would then reflect that of the stellar surface that is continuously heated by residual accretion and masks the interior temperature of the neutron star. We therefore searched for signatures of quiescent accretion in \exo.

X-ray monitoring with \swift\ has revealed X-ray flares in several quiescent neutron stars, e.g., \xte, Aql X-1, Cen X-4, \kstwee, \grs, and \sax\ \citep[e.g.,][]{bernardini2013,degenaar09_gc,degenaar2013_ks1741,degenaar2012_gc,fridriksson2011,wijnands2013,cotizelati2014}. During these X-ray flares the XRT count rate increased for several days by a factor of $\simeq$2 to even $>$10 for some of these sources. A corresponding hardening of the X-ray spectrum is observed and suggests that these flares are possibly caused by a spurt of low-level accretion. \exo\ was monitored with \swift\ roughly once per month for $\simeq$10~ks between 2008 and 2011. We did not detect any irregular X-ray variability or flaring events such as seen in other sources. Regular \swift\ monitoring has therefore not revealed any indications of ongoing low-level accretion in \exo. However, accretion flares appear to be short-lived events (lasting $\simeq$~days), and could therefore easily be missed \citep[e.g.,][]{degenaar09_gc,degenaar2013_ks1741,fridriksson2011,wijnands2013,cotizelati2014}. 

The first \chan\ and \xmm\ observations of \exo\ (obtained in 2008, within 2 months after the outburst end) both showed the presence of non-thermal emission, albeit contributing only $\lesssim$15\% to the total unabsorbed 0.5--10 keV flux \citep[whereas this is $>$50\% in some other neutron stars such as \saxamxp, \swiftpulsar, and \exoter; e.g.,][]{heinke2009,degenaar2012_1745,degenaar2012_amxp}. 
Optical spectroscopic and photometric observations performed shortly after the transition to quiescence hinted the presence of an accretion disk that could allow for continued accretion onto the neutron star \citep[][]{bassa09,hynes09}. However, optical spectroscopy and Doppler tomography performed one year later did not show evidence for an accretion disk any more \citep[][]{ratti2012}. Any contribution from non-thermal X-ray emission also remained low at this time (Section~\ref{subsec:evolution}). Finally, there are no dips or other features in the quiescent X-ray light curves that might evidence the presence of a residual accretion stream or remnant disk \citep[see also][]{diaztrigo2011}.

We conclude that there are no obvious signs of ongoing accretion in the quiescent state of \exo, particularly not $\gtrsim$1 yr after the outburst ended. Given the optical signatures of a quiescent accretion disk and the presence of non-thermal X-ray emission we cannot exclude, however, that matter was falling onto the neutron star shortly after the outburst appeared to have ended.

\acknowledgments
N.D. is supported by NASA through Hubble Postdoctoral Fellowship grant number HST-HF-51287.01-A from the Space Telescope Science Institute (STScI), which is operated by the Association of Universities for Research in Astronomy, Incorporated, under National Aeronautics and Space Administration (NASA) contract NAS 5-26555. Support for this work was provided by the NASA through Chandra Award No. GO3-14050X issued by the Chandra X-ray Observatory Center which is operated by the Smithsonian Astrophysical Observatory
for and on behalf of the NASA under contract NAS8-03060. A.C. is supported by an NSERC Discovery Grant and is an Associate Member of the CIFAR Cosmology and Gravity program. M.T.W. is supported by the Office of Naval Research. The authors are grateful to Neil Gehrels and the \swift\ planning team for approving and scheduling the \swift\ ToO observations of \exo. N.D., A.C., R.W., E.C., and E.B. acknowledge the hospitality of the International Space Science Institute in Bern, Switzerland, where part of this work was carried out. The authors are grateful to the anonymous referee and Joel Fridriksson for very useful comments. 

{\it Facilities:} \facility{{\it CXO} (ACIS), {\it XMM} (EPIC), {\it Swift} (XRT)}

\bibliographystyle{apj}

\end{document}